\documentstyle[epsfig]{mn2e}

\begin{document}

\title[Radio/X-ray Correlation]{On the Origin of the Universal Radio-X-Ray 
Luminosity Correlation In Black Hole Candidates}

\author[Robertson \& Leiter]{Stanley L. Robertson$^1$ and Darryl J. Leiter$^2$\\
$^1$Physics Dept.,Southwestern Oklahoma State University,
Weatherford, OK 73096, USA (stan.robertson@swosu.edu)\\
$^2$FSTC, Charlottesville, VA 22901, USA (dleiter@aol.com)\\}
\maketitle

\begin{abstract}
In previous work we found that the spectral state switch and other spectral 
properties of both neutron star (NS) and galactic black hole candidates 
(GBHC), in low mass x-ray binary systems could be explained by a magnetic 
propeller effect that requires an intrinsically magnetic central compact
object. In later work we showed that intrinsically 
magnetic GBHC could be easily accommodated by
general relativity in terms of magnetospheric eternally 
collapsing objects (MECO), with lifetimes greater than a 
Hubble time, and examined some of their spectral properties. 
In this work we show how a standard thin accretion disk 
and corona can interact with the central magnetic field 
in atoll class NS, and GBHC and active galactic nuclei (AGN) modeled as MECO,
to produce jets that emit radio through infrared luminosity $L_R$
that is correlated with mass and x-ray luminosity as 
$L_R \propto M^{0.75 - 0.92}L_x^{2/3}$
up to a mass scale invariant cutoff 
at the spectral state switch. Comparing the MECO-GBHC/AGN model to
observations, we find that the 
correlation exponent, the mass scale invariant cutoff, and the 
radio luminosity ratios of AGN, GBHC and atoll class NS are correctly
predicted, which strongly implies that GBHC and AGN have observable 
intrinsic magnetic moments and hence do not have event horizons.
\end{abstract}


\section{Introduction}
In earlier work (Robertson \& Leiter 2002, hereafter RL02) we extended analyses
of magnetic propeller effects (Campana et al. 1998, Zhang, Yu \& Zhang 1998)
of neutron stars (NS) in low mass x-ray binaries (LMXB) to the domain
of GBHC. From the luminosities at the low/high
spectral state transitions, accurate rates of spin were found for NS
and accurate quiescent luminosities were calculated for \textit{both} NS
and GBHC. The NS magnetic moments were found to be consistent with
$\sim 10^{8 - 9}$G magnetic fields, in good agreement with
those others have found (e.g. Bhattacharya 1995) from spin-down
rates for similarly spinning 200 - 600 Hz millisecond pulsars.
GBHC spins were found to be typically 10 - 50 Hz. Their magnetic moments of
$\sim 10^{29}$ gauss cm$^3$ are $\sim 200$ times larger than those
of `atoll' class NS (e.g. Burderi et al. 2002, DiSalvo \& Burderi 2003). 
The implied magnetic fields of GBHC are in good
agreement with fields of $\sim 10^8$G that have been found at the base of
the jets of GRS 1915+105 (Gliozzi, Bodo \& Ghisellini 1999,
Vadawale, Rao \& Chakrabarti 2001) and in the accretion disk of
Cygnus X-1 (Gnedin et al. 2003). At accretion disk inner radii
corresponding to the low/high spectral state switch, the magnetic fields of both
`atoll' class NS and GBHC are $\sim 5\times10^7$G, which may account for some
of their strong similarities (e.g. Yu et al. 2003, Tanaka \& Shibazaki 1996,
van der Klis 1994).

In later work (Leiter \& Robertson 2003, Robertson \& Leiter 2003,
hereafter RL03) we have described how the Einstein field equations
of General Relativity applied to compact plasmas with equipartition magnetic
fields permit the existence of magnetic, eternally collapsing objects (MECO)
that can have lifetimes in excess of a Hubble time. These highly redshifted,
faintly (as distantly observed in quiescence) radiating objects
can produce `ultrasoft' thermal spectral peaks and the magnetic propeller
effects found in RL02. Here we examine the accretion
disk - magnetosphere interaction and show how the magnetosphere
can drive jets. Our model should be applicable for any jet producing objects
with sufficiently large magnetic moments, whether T-Tauri stars or NS, or GBHC
and active galactic nuclei (AGN) modeled as MECO.
In this context, the scaling of the
magnetic moments of MECO with mass will be an important consideration.
As shown in RL03, the MECO is dominated by a photon-photon collision
generated pair plasma which is stabilized at high redshift deep inside the
photon orbit by an Eddington limit radiation pressure generated by
an equipartition magnetic field intrinsic to the MECO. The surface value of
the MECO intrinsic magnetic field is calculated
by equating the synchrotron generated photon pressure ($\propto B^4$) to the
gravitational force per unit area, which is proportional to the density.
Since the density is inversely proportional to the
square of the MECO mass, $M$, the internal magnetic field scales as $M^{-1/2}$
and the MECO magnetic moment, $\mu$, scales as $M^{-1/2}(2GM/c^2)^3
\propto M^{5/2}$. 

In the following, for NS, GBHC and AGN, we will assume the existence of a
gas pressure dominated, geometrically thin accretion disk (Shakura \&
Sunyaev 1973). For gas pressure dominance, it has been shown (e.g., Merloni
\& Fabian 2002) that the hard x-ray spectral tail and
reflection features of the low state spectrum can
be adequately explained by reprocessing of the soft thermal disk photons in
an accretion disk corona (ADC). The physical size of a corona is consistent
with limits found for the source of the power-law x-ray emissions of LMXB
(Church \& Baluci\'{n}ska-Church 2003).

It has been suggested, however, (Markoff, Falcke
\& Fender 2001, Falcke, K\"{o}rding \& Markoff 2003) that the power-law x-ray
emissions might originate in a jet. Flat or inverted spectrum synchrotron radio-
infrared emissions are generally believed to originate in jets and low state jets
have been resolved (Stirling et al. 2001) and studied over a wide range of
luminosity variation (Corbel et al. 2000, 2003). As a result of these
outflows, it has been pointed out (Fender, Gallo \& Jonker 2003) that
the low quiescent luminosities of GBHC cannot be taken as evidence of
advective accretion flows (ADAF) through event horizons and as noted
by Abramowicz, Kluzniak and Lasota (2002) there is presently no
other observational evidence of event horizons.

Whether or not the x-rays
originate in the jet, there is a strong coupling between x-ray and radio
emissions that must be related to the accretion flow and jet structure.
A universal low state radio / X-ray correlation ($L_R \propto L_x^{0.7}$)
(Gallo, Fender \& Pooley 2003) with a cutoff at the low/high state transition
(Fender et al. 1999, Tannenbaum et al. 1972, Corbel et al. 2003)
has been found for GBHC and NS ( Migliari et al. 2003).
A similar radio / x-ray correlation (Merloni, Heinz \& Di Matteo 2003, Falcke,
K\"{o}rding \& Markoff, 2003) and its suppression at the transition to the
high/soft state (Maccarone, Gallo \& Fender 2003) have been shown to hold 
for AGN as well. These radio / X-ray luminosity correlations
have been examined for scale invariant jets (Heinz \& Sunyaev 2003, hereafter
HS03), yielding constraints on the accretion processes. In the context of
HS03, Merloni, Heinz \& Di Matteo (2003) (Hereafter MHD03)
have examined their correlation for compatibility with
various accretion flow models and found better consistency with an
ADAF / jet model than with radiatively efficient disk / jet
or pure jet models.

An ADAF/jet model (Meier 2001) can also account for the low/high spectral state
transition as a transition from an ADAF to a standard thin disk.
It relies on a rapid black hole spin to provide energy to drive the jet.
The model predicts that stable high/soft states would not exist for AGN
more massive than $7 \times 10^4 M_\odot$ (Meier 2001) or, with more generous
allowance for hysteresis effects, $\sim 4\times 10^6 M_\odot$
(Maccarone, Gallo \& Fender 2003). The theoretical mass limit occurs
because the Eddington scaled luminosity at which a thin disk
(constrained to match a radiatively inefficient ADAF accretion rate)
becomes radiation dominated is mass dependent. Since
the high/soft state nevertheless appears to exist
in AGN more massive than $6 \times 10^7 M_\odot$, the ADAF transition
model cannot be regarded as established. Understanding the origin of the
mass limit error of the model remains an open question
(Maccarone, Gallo \& Fender 2003).

Black hole models that rely entirely on the jet to produce the power-law
x-ray emissions may have difficulties with constraints on the physical
size of a jet.  For dipping sources the size of the region of the low state
power-law production has been found (Church \& Baluci\'{n}ska-Church 2003)
to be $\leq 10^9$ cm. In addition, there is the enigma of
the size of the hard spectral producing region increasing
while the jet dies in the high state.
It is also unclear how black hole and NS
behaviours could be so similar with the magnetic fields of even the weakly
magnetized atoll class NS being capable of disrupting the inner accretion disk.
On the other hand, we will show that our MECO model, with a radiatively
efficient disk, will provide a superior fit to the radio / X-ray correlations
and provide a mass scale invariant cutoff at the high/soft state transition
while permitting radio-infrared 
and some of the x-ray luminosity to originate in a jet.

\section{The Disk - Intrinsic Magnetic Moment Interaction}
In the magnetic propeller model, the inner disk and magnetosphere radius, $r_m$,
determines the spectral state. Very low to quiescent states correspond
to an inner accretion disk radius outside the light cylinder.
In the low/hard/radio-loud/jet-producing state of
the active propeller regime, the inner disk
radius lies between light cylinder and Keplerian co-rotation
radii. Most, and perhaps all, of the accretion flow is ejected in
the low/hard state. The high/soft state corresponds to an
inner disk inside the co-rotation radius with the flow of accreting matter
able to reach the central object where it produces an ultrasoft thermal
spectral component. The cooling of the accretion
disk corona and the former base of the jet by the soft photons also
contributes to a softening of the x-ray spectrum. The whole complex
of spectral state switch phenomena is related to the cessation or
regeneration of magnetospherically driven outflow and presence or absence
of dominant soft emissions from a central source.

The inner disk temperature is generally high enough
to produce a very diamagnetic plasma at the magnetopause. Surface currents
on the inner disk distort the magnetopause and they
also substantially shield the trailing disk such that the
region of strong disk-magnetosphere interaction is mostly confined to a ring
or torus, of width $\delta r$ and half height $H$. This shielding leaves most
of the disk under the influence of its own internal shear dynamo fields,
(e.g. Balbus \& Hawley 1998, Balbus 2003). At the inner disk
radius the magnetic field of
the central MECO is much stronger than the shear dynamo field generated
within the inner accretion disk. In MHD approximation, the force density
on the inner ring is $F_v = (\nabla \times B) \times B / 4\pi$.
For simplicity, we assume coincident magnetic and spin axes of the central
object and take this axis as the $z$ axis of cylindrical coordinates
$(r,\phi,z)$.
 
The magnetic torque per unit volume of plasma in the inner ring of
the disk that is threaded by the intrinsic magnetic field of the
central object, can be approximated by $\tau_v=rF_{v\phi} =
r \frac{B_z}{4\pi} \frac{\partial B_{\phi}}{\partial z}
\sim r \frac{B_zB_{\phi}}{4\pi H}$, where $B_{\phi}$
is the average azimuthal magnetic field component. 
We stress that $B_{\phi}$, as used here,
is an average toroidal magnetic field component. The toroidal
component likely varies episodically between reconnection events
(Goodson \& Winglee 1999, Matt et al. 2002, 
Kato, Hayashi \& Matsumato 2004, Uzdensky 2002).

The average flow of disk angular momentum entering the inner ring
is $\dot{M}r v_k$, where $\dot{M}$ is mass accretion rate and $v_k$ is the
Keplerian speed in the disk. This angular momentum
must be extracted by the magnetic torque, $\tau$, hence:
\begin{equation}
\tau = \dot{M}r v_k = r\frac{B_zB_{\phi}}{4\pi H}(4\pi r H \delta r).
\end{equation}
In order to proceed further, we assume that $B_{\phi} = \lambda B_z$,
$B_z=\mu/r^3$, and use $v_k=\sqrt{GM/r}$, where $\lambda$ is a constant,
presumed to be of order unity, $\mu$ is the magnetic dipole moment of
the central object $M$, its mass, and $G$, the Newtonian gravitational
force constant. With these assumptions we obtain
\begin{equation}
\dot{M} = (\frac{\lambda \delta r}{r}) \frac {\mu^2}{r^5 \omega_k}
\end{equation}
where $\omega_k = v_k/r$ and the magnetopause radius, $r_m$ is given by
\begin{equation}
r_m = (\frac{\lambda \delta r}{r})^{2/7}(\frac{\mu^4}{GM\dot{M}^2})^{1/7}
\end{equation}
We scale the accretion rate to that needed to produce luminosity at the
Eddington limit for a central object of mass $M$, and define
\begin{equation}
\dot{m}= \frac{\dot{M}}{\dot{M}_{Edd}} ~\propto~ \frac{\dot{M}}{M}
\end{equation}
and using $r_g=GM/c^2$ and eq. 3, we define
\begin{equation}
\chi = r/r_g ~\propto~ \dot{m}^{-2/7}.
\end{equation}

In order to estimate the size of the boundary region, $(\delta r /r)$,
we normalize this disk-magnetosphere model to an average atoll NS
(Table 1, RL02) of mass $M = 1.4 M_{\odot}$. The average rate of spin
is $\sim 450$ Hz, the co-rotation radius is $\sim 30$ km, and the maximum
luminosity for the low state is $GM\dot{M}/2r ~\sim 2 \times 10^{36}$
erg s$^{-1}$ From this we find $\dot{M} = 6.4 \times 10^{16} g s^{-1}$.
Then for an average magnetic moment of $\sim 1.5 \times 10^{27}$ gauss cm$^3$,
we find that ($\frac{\lambda \delta r}{r})^{2/7} \sim 0.3$. Thus
$\xi = \frac{\lambda \delta r}{r} \sim 0.015$; i.e., the boundary region is
suitably small, though likely larger than the scale height of the trailing disk.
For later convenience, we define parameters $\beta$ and $\xi$ as
\begin{equation}
\beta = \mu/M^3 ~~~~~~ \xi = \lambda \delta r /r
\end{equation}
Then in terms of the variables defined so far, we can express the (reprocessed)
disk luminosity as
\begin{equation}
L_d = \frac{GM\dot{M}}{2r}=\xi \frac{\sqrt{GM} \mu^2}{2 r^{9/2}}=\frac{\xi \mu^2 \omega_k}{2r^3}
    ~\propto~ \beta^2 M^2 \dot{m}^{9/7}
\end{equation}
At the co-rotation radius we reach the maximum luminosity, $L_c$,
of the low/hard state, with Keplerian angular speed $\omega_k = \omega_s$, the
angular speed of the magnetosphere, and
$r_m=(GM/\omega_s^2)^{1/3}$. Thus
\begin{equation}
L_c=\xi\frac{\mu^2 \omega_s^3}{2GM}
\end{equation}
Two additional quantities needed for the analysis of the flow into the
base of a jet are the scaling parameters for the poloidal magnetic field
and the inner disk density. The magnetic field at the base of the jet is
simply
\begin{equation}
B_m = \frac{\mu}{r_m^3} ~\propto ~ \frac{\beta}{\chi_m^3}
\end{equation}
For $\rho$, the density in the disk, we assume a standard `alpha' disk,
for which $\dot{M} = \rho 4 \pi r H v_r$. The radial inflow speed, $v_r$
is proportional to $v_s H/r$, where $v_s$ is sound speed in the disk.
Using $H/r \propto v_s/v_k$, taking $v_s^2 \propto B^2/\rho$ and solving
the mass flow rate equation for $\rho$ yields:
\begin{equation}
\rho_m ~\propto~ \frac{\mu^2}{M r_m^5} ~\propto~ \frac{\beta^2}{\chi_m^5}
\end{equation}

\subsection{Mass Ejection and Radio Emission}
The radio luminosity of a jet is a function of the rate at which the
magnetosphere can do work on the inner ring of the disk. This depends 
on the relative speed between the magnetosphere and the
inner disk; i.e., $\dot{E} =\tau (\omega_s - \omega_k)$, or
\begin{equation}
\dot{E} = \xi\frac{\mu^2 \omega_s (1 - \frac{\omega_k}{\omega_s})}{r^3}
    ~\propto~ \mu^2 M^{-3}\dot{m}^{6/7}\omega_s(1 - \frac{\omega_k}{\omega_s})
\end{equation}

Disk mass, spiraling in quasi-Keplerian orbits from negligible speed
at radial infinity must regain at least as much energy as was radiated
away in order to escape. For this to be provided by the magnetosphere
requires $\dot{E} \geq GM\dot{M}/2r$, from which $\omega_k \leq 2\omega_s/3$.
Thus the magnetosphere alone is incapable of completely ejecting all
of the accreting matter once the inner disk reaches this limit and
the radio luminosity will be commensurately reduced and ultimately cut off.
\footnote{For very rapid inner disk transit through the co-rotation radius,
fast relative motion between inner disk and magnetosphere can heat the inner
disk plasma and strong bursts of radiation pressure from the central object
may drive large outflows while an extended jet structure is still
largely intact. This process has been calculated for inner disk radii inside
the marginally stable orbit (Chou \& Tajima 1999) using pressures
and poloidal magnetic fields of unspecified origins. A MECO is obviously
capable of suppling both the field and a radiation pressure. The hysteresis of
the low/high and high/low state transitions may be associated with
the need for the inner disk to be completely beyond the corotation
radius before a jet can be regenerated after it has subsided.}

The radio flux, $F_{\nu}$, of jet sources has a power law dependence
on frequency of the form
\begin{equation}
F_{\nu}~ \propto~ \nu^{-\alpha}
\end{equation}
The spectral energy distributions of GBHC and AGN in radio-infrared show very
little, if any, evolution in the low state during outbursts; i.e., $\alpha$
is essentially constant ($\alpha \approx -0.5$, radio; $-0.15$, IR, see e.g.,
Chaty et al. 2003). To determine the dependence of radio flux on $\mu$,
$M$ and $\dot{m}$, we use the model and methods
of HS03. Their  analysis was based on
a radiation transfer equation (Rybicki \& Lightman 1979) which gives the
radio flux from a jet viewed at right angle to the jet axis as
\begin{equation}
F_{\nu}~\propto~ \int R(z)^2 j_{\nu}(z)[1-\exp{(-\tau_{\nu}(z))}]/\tau_{\nu}(z)dz
\end{equation}
Here $z$ is a coordinate along the conical jet axis of symmetry,
$R(z)$ is the radius of the jet, $j_{\nu}(z)$ is the optically thin synchrotron
emissivity for a power law distribution
of electrons over energy and $\tau_{\nu}(z)$ is the optical depth for a viewing
angle perpendicular to the jet axis. Noting that $\tau_{\nu}(z)$ becomes huge
below the shock at the base of the jet, the integral can be taken from
$z \approx 0$
to $z \rightarrow \infty$. Both $j_{\nu}(z)$ and $\tau_{\nu}(z)$ depend on
the magnetic field and density distributions along the jet. We assume that
the magnetic field will be proportional to $B_m$ of eq. (9). The
density after passage through the jet shock will remain proportional to
$\rho_m$ of eq. (10). Thus we assume that $B(z)=\beta f(z)/\chi_m^3$ and
$\rho(z) = g(z)\beta^2/\chi_m^5$, where f(z) and g(z) are distribution functions
along the jet.

In order to evaluate the integral for
$F_{\nu}$, it is helpful to scale $z$ and $R(z)$ to match the disk radius
at the base of the jet nozzle. For this purpose, we define
variables scaled the same as $\chi$; $\zeta = z \dot{m}^{-2/7}/r_g$
and $R_{\zeta}(\zeta)= R(z)\dot{m}^{-2/7}/r_g$.
With this scaling, $R_{\zeta}$ can automatically always
match $\chi_m$ at the base of the jet and:
\begin{equation}
F_{\nu} \propto M^3\dot{m}^{6/7}\int_0^{\infty} 
    R_{\zeta}(\zeta)^2 j_{\nu}(\zeta)[1-\exp{(-\tau_{\nu}(\zeta))}]/\tau_{\nu}(\zeta)d\zeta
\end{equation}
The integral above has magnetic field and density dependence
only via $\beta = \mu/M^3$ in Equations (9) and (10).
For given inner disk radius in units of $r_g$, having $B_m$ determined 
predominantly by the central object represents
a case that was not considered in HS03. Nevertheless, using their method
(and with notation adapted from their eq. (8) to the present case,
$\phi_B= \beta$ and $\phi_C=\beta^2$) we obtain $\alpha$ from a
differentiation of the logarithm of the integral with respect to
$\ln(\nu)$. A second differentiation of $\ln{(\alpha)}$ with respect to
$\ln{(\dot{m})}$ yields a zero because the MECO magnetic field, 
and thus $\beta$, is independent of $\dot{m}$, thus
assuring that there is no low state spectral evolution as $\dot{m}$ changes.
Further, following HS03, we assume scale invariance of the jet morphology.
For given $\chi_m$, the integral is invariant with respect to $\dot{m}$.

The scaling of $B$
and $\rho$ satisfies the conditions for applicability of the method used
by HS03 to obtain their eq. (10a). Then by similar method we
obtain the dependence of $F_\nu$ on M and $F_{\nu} \propto \beta^q$, where
\begin{equation}
q=\frac{13+2p+\alpha(p+6)}{p+4}.
\end{equation}
Taking the canonical value of $p=2$, we obtain $q=(17+8\alpha)/6$ and 
for the accretion disk-intrinsic magnetic moment interaction and spectral index
described by equations (1) through (14) we find
\begin{equation}
F_{\nu} \propto M^{(17+2\alpha)/6} \dot{m}^{6/7} \beta^{(17+8\alpha)/6}\nu^{-\alpha}
\end{equation}
If $\beta \propto M^{-1/2}$, as would apply for MECO AGN/GBHC this 
recovers the HS03 dependence of $F_{\nu} \propto M^{(17/12-\alpha /3)}$
from their eq. (10a), but for strict scale invariance of the integral,
there is no further dependence on $\dot{m}$ here. This differs from the 
$\dot{m}$ dependence found by HS03 because the dominant magnetic field
of the jet originates in the MECO rather than being generated in the accretion
flow of the disk.

With $\mu$ in eq. (11) written in terms of $\beta$, a comparison
with eq. (16) shows the radio flux to be proportional to $\dot{E}$.
Thus we can take the integrated radio flux as luminosity, $L_R$, to be given by
\begin{equation}
L_R = C' \dot{E} = C_o M^{(2\alpha -1)/6}\beta^{(5+8\alpha)/6}\dot{E}/\omega_s
\end{equation}
where $C_o$ is a constant dependent on the radio bandwidth.

As noted by HS03, there will also be optically thin x-ray emission from the
jet. In this case, taking $\tau_{\nu} << 1$ in eq. 14, we obtain what is
essentially an integral over the jet source volume for the optically
thin x-ray emission of the jet. Since $F_{\nu,x}$ depends
on $\dot{m}$ in the same way as before,
both radio-infrared and the jet part of the x-ray fluxes are proportional to
$\dot{E}$. While the base of the jet contributes to the x-ray flux, its
radiating volume is likely much smaller than that of the ADC, which produces
most of the x-ray flux. The cutoff of this part of the x-ray flux and the
onset of soft thermal emissions as the inner disk radius pushes inside
corotation and the accretion flow reaches the central object
marks the spectral state transition. The ADC actually grows in the high
state (while producing a declining fraction of the x-ray luminosity)
as it is cooled by photons from the central object.

Finally, we note that the degree of collimation of a jet actually appears to
depend on the scale height and pressure of the corona (Kato, Minishige \& Shibata
2004), but $F_\nu$ can still be calculated for a largely uncollimated outflow;
for example, a large angle flow spreading out from the inner rings of the disk.
In this case, we would obtain
an integral similar to eq. 14. with $R_\zeta^2 d\zeta$ replaced by a column
length parallel to the line of sight looking into the plasma and
integrated over the area of the flow, projected perpendicular to the line of
sight. Though there would be differences of numerical factors depending on
viewing angle, scale invariance for given $r_m/r_g = \chi_m$ would still require
all coordinates to be scaled in terms of $r_g$ and $\dot{m}^{2/7}$
in the same way as in eq. 14 and we would still obtain $F_\nu \propto \dot{E}$.
Thus the scaling results we have obtained for magnetospherically
driven outflows are very robust, even though there may be
considerable uncertainty about the geometric details of the flow.

\subsection{Radio - X-ray Correlation}
Since $\dot{E} \propto r^{-3}$ and $L_d \propto r^{-9/2}$, it is apparent
that we should expect radio luminosity, $L_R \propto L_d^{2/3}$.
In particular we find
\begin{equation}
L_R = C(M,\beta, \omega_s)2L_c^{1/3} L_d^{2/3}(1-\omega_k/\omega_s)
\end{equation}
where
\begin{equation}
C(M,\beta, \omega_s)=C_o M^{\frac{(2\alpha -1)}{6}}\beta^{\frac{(5+8\alpha)}{6}}/\omega_s
\end{equation}
Strictly speaking, $L_d$ should be the bolometric luminosity of the disk,
however, the x-ray luminosity over a large energy band is a very substantial
fraction of the disk luminosity. To compare with the correlation exponent of
2/3 obtained here, recent studies, including noisy data for both GBHC and AGN
have yielded $0.71 \pm 0.01$ (Gallo, Fender \& Pooley 2003),
0.72 (Markoff et al. 2003, Falcke, K\"{o}rding \& Markoff 2003), $0.60 \pm 0.11$
(MHD03) and $0.64 \pm 0.09$ (Maccarone, Gallo \&
Fender 2003). For $\alpha$ in the range 0, -0.5, $\beta \propto M^{-1/2}$,
$\omega_s \propto M^{-1}$ and $L_c \propto M$, the MECO model yields
$C(M) \propto M^{(9-4\alpha)/12}$ and
\begin{equation}
log L_R = (2/3)log L_x + (0.75 - 0.92) log M + const.
\end{equation} which is a better fit to the "fundamental plane" of
MHD03 than any of the ADAF, disk/corona or
disk/jet models they considered (see their Figure 5 for a $\chi^2$
density plot).
In terms of Eddington scaled luminosities $L_{R,Edd} = L_R/L_{Edd}$
and $L_{d,Edd} = L_d/L_{Edd}$ eq. (18) can be written as
\begin{equation}
L_{R,Edd} = C(M,\beta, \omega_s) 2L_{c,Edd}^{1/3}
        L_{x,Edd}^{2/3}(1-\omega_k/\omega_s)
\end{equation}

Another outstanding physical property of the MECO model for GBHC and AGN
is that the Eddington scaled x-ray luminosity, $L_x/L_{Edd}$, which is
$\propto L_c/M$ at the spectral state switch, is mass scale invariant. With
$\mu$ scaling as $M^{2.5}$ and $\omega_s$ as $M^{-1}$, eq. (8) shows that
$L_c/M$ is constant. $L_c/L_{Edd} \approx 0.02$ has been found for GBHC and
AGN (Ghisellini \& Celotti 2001, Maccarone, Gallo \& Fender 2003), and
remarkably, a similar value is found for atoll class NS (Tanaka \& Shibazaki
1996, RL02). With this common ratio, the Eddington scaled radio luminosities
of MECO objects will correlate with x-ray emissions at the low/high spectral
state switch in proportion to $C(M,\beta, \omega_s)$. For $\alpha$ in the 
range 0, -0.5, $C(M, \beta, \omega_s) \propto M^{0.42 - 0.58}$ 
for MECO AGN/GBHC, in good agreement with the MHD03 correlation.

Even though they are not MECO the behavior of NS x-ray binaries can also be 
described by our model since they contain a central object with an intrinsic 
magnetic moment. According to eq. 18, the ratio of peak
radio luminosity to $L_c$ is also just proportional to $C(M,\beta,\omega_s)
\sim \mu^b/(M^{3b-a} \omega_s)$, where $a=(2\alpha-1)/6$
and $b=(5+8\alpha)/6$. Using average values of
$\mu, M$ and $\omega_s$ for the limited sample from Table 1 of RL02,
and with $\alpha$ in the range 0, -0.5, we predict that GBHC should
have peak radio / x-ray luminosity ratios that are 10 - 13 times
larger than for atoll NS. Fender and Kuulkers (2001)
compared the ratio of radio to x-ray luminosities at the radio peak for
GBHC and NS. Omitting two GBHC and one
NS that are abnormally radio loud the comparison between GBHC and atoll class
NS ratio averages is $\sim 31/2.4 = 13$, from data in their Table 1.

Both $L_c$ and peak
radio emissions for Z class NS are larger because their magnetic moments are
about 10X larger than those of atolls, but $L_c$ increases by less than $10^2$,
because on average they likely spin more slowly than atolls. These Z sources and
abnormally loud radio sources, such as Cygnus X-3 can be explained with the
magnetic model. It has been suggested that Cygnus X-3 is a NS (Brazier
et al. 1990, Mitra 1998). Based on its apparent 79 Hz spin and spin down rate,
it should have a magnetic moment about
13X that of an average MECO-GBHC and spin about 2.6X faster than an average 
MECO-GBHC (RL02). Given an adequate accretion supply, its low state steady radio
emissions (not bubble events) could be $13^2 2.6^3 \sim 3000$ times stronger
than an average MECO-GBHC. 

\begin{figure}
\epsfig{figure=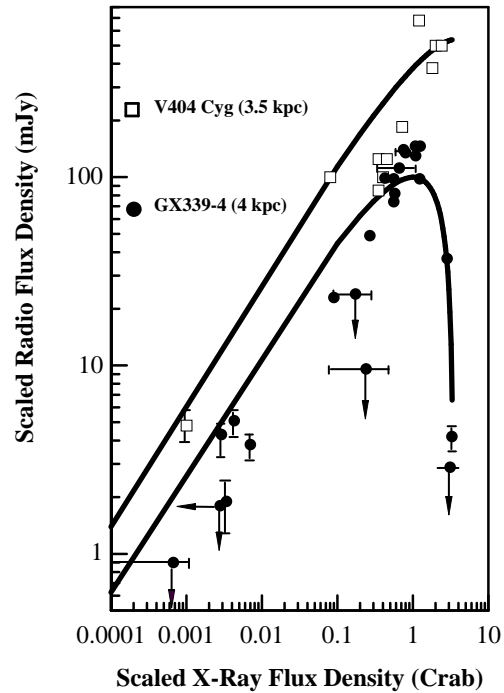,angle=0,width=8cm}
\caption{Radio against X-Ray flux density, scaled to a distance of
1 kpc and absorption corrected for V404 Cygni and GX 339-4. Error limits
for V404 Cygni are approximately the box size, except where otherwise indicated.
Lines denote fits of eq. 22 to the data. The cutoff for
V404 Cygni is based on a previously identified x-ray spectral state switch
(RL02). Data were kindly provided by Elena Gallo.}
\end{figure}

If we let $x=L_d/L_c$, then for $x < 1$, corresponding to the low state,
eq. (18) takes the form:
\begin{equation}
L_R =C(M,\beta, \omega_s) 2L_c(x^{2/3} - x)
\end{equation}
Neglecting $C(M,\beta, \omega_s)$, the
function has a maximum value of $(0.3 L_c)$ at $x=0.3$. Scaling the
function for $L_c$ at the cutoff or $0.3 L_c$ at the maximum is quite easy.
An `eyeball' fit to the data of Figure 1 provides a sense of the sharpness 
of the cutoff associated with $L_c$. Since $\omega_s$ is
known for several NS from burst oscillations, eq. (8) has been used (RL02) to
constrain their magnetic moments.

$L_c$, in  eqs. (8) and (22), is an individual cutoff luminosity that depends
on $\omega_s$, which varies, depending upon the
average accretion rate produced by a binary companion or an AGN environment.
Consequently the mass scale invariant cutoff ratio $L_c/ L_{Edd}$ has a random
variability, but apparently within only a narrow range.
At any epoch, it is likely that most GBHC and NS are near spin
equilibrium; being neither spun up nor spun down by accretion on
an average over a few outburst cycles. 
In high states they can be spun up by accretion while in 
low states they are spun down by magnetospherically driven outflows.
If the magnetic fields of NS weaken with age, for whatever reason,
then they would be spun up commensurately by accretion
disks that can then more easily penetrate to smaller radii and drive the
magnetosphere to higher Keplerian angular speeds. On the other hand 
the highly redshifted, slow collapse of a MECO stabilizes
its intrinsic equipartition magnetic field and leads us to expect very 
little field decay for the MECO-GBHC. Since the MECO intrinsic
magnetic field is determined solely by its mass, the correct mass scaling
found here for MECO-GBHC/AGN, which was based on $\omega_s \propto M^{-1}$
without additional multipliers, suggests that MECO/AGN
most likely are in a state of slow spin equilibrium.

\section{Conclusions}
In a previous paper (RL02) we found that the spectral state switch and
other spectral properties of low mass x-ray binaries, including 
both NS and GBHC, could be explained by a magnetic 
propeller effect that requires an intrinsically magnetized 
central object. Subsequently (RL03) we applied the Einstein field
equations of General Relativity to the case of a highly
compact, Eddington limited, pair dominated plasma with an
intrinsic equipartition magnetic field. We found that the Einstein equations 
permit the existence of intrinsically magnetic, highly red shifted, 
extremely long lived, collapsing, radiating MECO objects 
that can produce the required propeller effects. 
In addition to accounting for the strong spectral similarities of NS and GBHC,  
the magnetosphere-accretion disk interaction associated with the MECO 
model has provided explanations for radio / x-ray luminosity correlations, 
the mass scale invariant spectral 
state switch phenomenon with its suppression of the radio jet 
outflow in the high/ soft state, the "ultrasoft" thermal peak
and hard spectral tail of the high state, and, finally, the quiescient
luminosities described as spin-down driven radiations.

In conclusion, we have shown here how a standard, thin, gas pressure
dominated accretion disk and corona can interact with the central 
intrinsic magnetic moments of MECO-GBHC/AGN and NS in x-ray biniaries
to drive low state jets. In the case of the MECO-GBHC/AGN the
radio-infrared emissions of the jets have been found to correlate 
with the x-ray luminosity up to a mass scale invariant cutoff $L_c / L_{Edd}$
at the spectral state switch. In this context we obtained radio-infrared
luminosities for MECO that vary as $M^{0.75-0.92}L_x^{2/3}$, consistent with
observations of GBHC and AGN, and correctly predicted the observed relative
radio luminosities of NS, GBHC, and AGN. While much detailed work remains to
be done, the successful comparison of the MECO model predictions with 
observations strongly suggests that GBHC and AGN may have observable
intrinsic magnetic moments anchored within them and hence they
do not have event horizons.\\

\noindent {\bf Acknowledgements}\\
We thank the anonymous referee for many comments and suggestions that have
substantially improved this paper.
We thank Elena Gallo for providing data for Figure 1.
Useful information has been generously provided by Mike Church, Heino Falcke
and Thomas Maccarone. We are very grateful to Abhas Mitra for many
helpful discussions of gravitational collapse and pertinent
astrophysical observations.

\end{document}